# Manipulation of Dirac band curvature and momentum-dependent *g*-factor in a kagome magnet YMn$_6$Sn$_6$

Hong Li[1], He Zhao[1], Kun Jiang[1], Qi Wang[2], Qiangwei Yin[2], Ning-Ning Zhao[2], Kai Liu[2], Ziqiang Wang[1,*], Hechang Lei[2,*] and Ilija Zeljkovic[1,*]

[1] *Department of Physics, Boston College, 140 Commonwealth Ave, Chestnut Hill, MA 02467*

[2] *Department of Physics and Beijing Key Laboratory of Opto-electronic Functional Materials & Micro-nano Devices, Renmin University of China, Beijing 100872, China*

[*]Correspondence: hlei@ruc.edu.cn , wangzi@bc.edu and ilija.zeljkovic@bc.edu

## Abstract

**The Zeeman effect describes the energy change of an atomic quantum state in magnetic field. The magnitude and the direction of this change depend on the dimensionless Landé *g*-factor. In quantum solids, the response of the Bloch electron states to the magnetic field also exhibits the Zeeman effect with an effective g-factor that was theoretically predicted to be dependent on the momentum. While typically negligible in many ordinary solids, the momentum-dependent variation of the g-factor is theorized to be substantially enhanced in many topological and magnetic systems. However, the momentum-dependence of the g-factor is notoriously difficult to extract and it is yet to be directly experimentally measured. In this work, we report the experimental discovery of a strongly momentum-dependent *g*-factor in a kagome magnet YMn$_6$Sn$_6$. Using spectroscopic-imaging scanning tunneling microscopy, we map the evolution of a massive Dirac band in the vicinity of the Fermi level as a function of magnetic field. We find that electronic states at different lattice momenta exhibit markedly different Zeeman energy shifts, giving rise to an anomalous *g*-factor that peaks around the Dirac point. Our work provides the first momentum-resolved visualization of Dirac band curvature manipulation by magnetic field, which should in principle be highly relevant to other topological kagome magnets.**

The Zeeman effect, first observed as the atomic spectral line splitting under an applied magnetic field, has found application across different branches of physics. The effect is rooted in the linear coupling between the total magnetic moment and the external magnetic field $\vec{B}$ detectable via the field-induced energy shift $-g(\mu_B/\hbar)\vec{J}\cdot\vec{B}$, where $\vec{J}$ is the spin-orbit angular moment of the atomic states and $g$ is a dimensionless number called Landé $g$-factor. Bloch electrons in quantum solids also exhibit the Zeeman coupling to the magnetic field set by an effective $g$-factor that takes on a more complex functional form. In general, the $g$-factor can vary with the overall band structure, spin-orbit coupling, carrier concentration, and exchange and correlation effects [1–5], and it is fully descried by a matrix that is a function of the crystal momentum $k$ [2]. While the theoretical framework behind the $g$-factor momentum dependence was established around the middle of the last century [1,2,6,7], the variations of the $g$-factor with momentum in ordinary metals and semiconductors are largely negligible.

Certain materials however are theorized to exhibit a strongly momentum-dependent effective $g$-factor, which can lead to substantial field-induced modifications to the electronic band structure. In graphene and topological materials, such as the twisted bilayer graphene, topological and Chern insulators, and Dirac and Weyl semimetal, $g$-factor can in principle be enhanced by the nontrivial topological band dispersion through momentum-dependent orbital magnetizations [8–14]. In magnetic and/or spin-textured systems, external magnetic field may further give rise to unusual Zeeman energy shifts and strong renormalization of the bands, enabling spin-direction-driven field-sensitive tuning of the electronic band structure [15–18] and topological phase transitions [18]. However, a direct momentum-resolved insight into the field-driven changes to the electronic band structure by experimentally measuring the shift of the quantum states in an external magnetic field has been difficult to achieve. Here we report the experimental discovery that the effective electron $g$-factor associated with a massive Dirac band, determined directly by measuring the Zeeman energy shift in a kagome magnet YMn$_6$Sn$_6$, is strongly momentum-dependent. Using spectroscopic-imaging scanning tunneling microscopy, we map a massive Dirac dispersion crossing the Fermi level as a function of magnetic field. We find that the energy of each state within the Dirac band shifts linearly with applied field. By charting the field-induced shift of each electronic state within the band, we map a strong $k$-dependent renormalization of the band curvature due to Zeeman coupling, which can be explained by a $k$-dependent electron $g$-factor.

Layered crystalline materials composed of atoms arranged on a lattice of corner-sharing triangles (kagome lattice) have emerged as a rich materials platform to study electronic correlations and non-trivial topology [19–21]. A simple tight-binding calculation of an isolated kagome plane, taking into account nearest-neighbor hopping, reveals that the system hosts Dirac points near the edge of the Brillouin zone and a dispersionless flat band [20,21]. Motivated by theoretical predictions of new electronic states, such as a fractional quantum Hall state without external field [22–24], spin liquid phases [25,26], tunable Weyl nodes [18] and a Wigner crystal [27], exotic phenomena have been intensely sought after and realized in several kagome compounds [11,28–40].

YMn$_6$Sn$_6$ is a layered material from a rare-earth (*Re*) family of kagome magnets *Re*Mn$_6$Sn$_6$ with an in-plane hexagonal structure ($a$ = $b$ = 0.554 nm, $c$ = 0.901 nm) [41]. It goes through an antiferromagnetic transition at $T_N \sim$ 350 K [41], which doubles the unit cell shown in Fig. 1a along the *c*-axis. As the temperature decreases further, the spin structure of YMn$_6$Sn$_6$ becomes helical [41–44]. A key ingredient of the crystal structure is the kagome lattice of Mn atoms, characterized by a network of corner-sharing Mn triangles, sandwiched between alternating stacks of Sn$^1$-Sn$^2$-Sn$^1$ and YSn$^3$ layers (Fig. 1a). We cleaved and investigated the surface of YMn$_6$Sn$_6$ single crystals grown by the flux method as described in Ref. [41] (Methods). STM topographs typically show surface terminations parallel to the *ab*-plane (Fig. 1c), although we occasionally found small patches of other crystalline facets as well (Fig. S3). We classify the *ab*-plane termination layers based on the morphology of the STM topographs, topographic terrace heights and average differential conductance (d$I$/d$V$) spectra (Fig. 1d-k, Supplementary Information 1). These experimental identifications are further verified by the simulated STM images from the first-principles calculations (Fig. S2).

We focus on the electronic structure of the Mn kagome surface. Average d$I$/d$V$ spectra show a sharp peak near Fermi level centered at $E_p \sim$ - 7.0 ± 0.1 meV (Fig. 1g and Fig. 2). This peak is spatially homogeneous, which rules out localized bound states tied to crystal defects (Fig. S4). To gain further insight, we perform spectroscopic measurements in magnetic field *B* applied parallel to the *c*-axis (perpendicular to the sample surface). We track an identical 40 nm square field of view at different fields, and compare average d$I$/d$V$ spectra (Fig. 2a,b). Aside from the spectral peak at $E_p$, we observe no additional peaks that could be interpreted as Landau levels up to the highest field used in our experiments (Fig. S7). If the peak at $E_p$ was associated with a doubly spin degenerate band, we would expect the two spin branches to shift in opposite directions due to spin Zeeman coupling; correspondingly, the spectral peak in d$I$/d$V$ would split. In contrast to this, we find that the peak does not split (nor broaden) up to the maximum *B* applied of 8 T (Fig. 2a,b). This points towards a singly spin polarized band. For a fixed spin orientation regardless of external *B*, we would expect a diverging shift of $E_p$ for the field applied parallel and antiparallel to *c*-axis. This possibility is again contradictory to our observations, as $E_p$ shifts in the same direction even if the direction of *B* is reversed (Fig. 2). Such a shift is consistent with a minority spin state in an electronic state polarized by the external field itself. Conventional Zeeman spin coupling characterized by electron *g*-factor $g_e \approx 2$ would dictate that the state shifts to lower energy with velocity $\frac{1}{2} g_e \cdot \mu_B \approx$ 0.058 meV/T ($\mu_B$ is one Bohr magneton). In contrast, we find that the state moves to a higher energy by five-fold the velocity of 0.31 meV/T (Fig. 2c). This amounts to a magnetic moment of -5.34$\mu_B$ or a large effective *g*-factor of $g_p$ = - 10.7. Note that the "dip" spectral feature (green arrows in Fig. 2a) shifts to higher energies faster than that of the adjacent peak $E_p$, which indicates a complex response of the electronic structure to the magnetic field in this system reaching beyond a rigid band shift.

A more detailed momentum-space picture of the low-energy electronic band structure can be obtained from spectroscopic-imaging STM (SI-STM). This method, also commonly referred to as quasiparticle interference (QPI) imaging, relies on elastic scattering and interference of electrons detectable as static modulations in d$I$/d$V$(**r**,*V*) maps. Let us first focus on the SI-STM measurement at zero magnetic field. Real-space d$I$/d$V$(**r**,*V*) maps show intense spatial modulations (Fig. 3b, top row), with nearly isotropic Fourier space wave vector $\boldsymbol{q_1}$ that rapidly evolves with energy (Fig. 3b, bottom row), disappearing around -15 meV (Fig. 3c, Fig. S6). From the QPI data, we extract the corresponding dispersion of the band crossing the Fermi level (Fig. 3c). Note that dispersion shows linear energy-momentum dependence at higher energies, but deviates from the linear dependence near the band bottom and acquires substantial curvature (outlined by dashed line in Fig. 3c). By comparing the Fermi velocity and the morphology of the QPI signal to the low-energy electronic band structure near the Fermi level measured recently by angle-resolved photoemission spectroscopy (ARPES) [45] (Fig. 3d), we can attribute the QPI dispersion to gapped Dirac cones at K, where the gap is a consequence of spin-orbit coupling. The sharp density-of-states peak at $E_p$ in Fig. 2 is then likely associated with the van Hove saddle point at M of a different band, located only a few millielectronVolts above the Dirac band bottom (Fig. 3d, Fig. 4a,b). The partial flat band observed by ARPES in the second Brillouin zone is located about 20 meV below the Dirac point [45], and thus unlikely to be associated with the sharp spectral feature at $E_p$. We note that the lower half of the Dirac cone cannot be unambiguously discerned in our data, possibly due to a decreased quasiparticle lifetime away from the Fermi energy. This is similar to the markedly weaker electronic signature of the upper half of the Dirac cone in related material $TbMn_6Sn_6$ [46], which is also positioned further away from zero energy.

Next we apply an external magnetic field of up to 8 T in both *c* and *-c* directions. Similar to the behavior of the spectral peak at $E_p$ (Fig. 2), the band dispersion does not split and shifts to higher energies independent of the direction of the magnetic field, thus pointing towards a magnetization polarized origin of a minority band (Fig. 4a-d). The quantitative field-induced energy shift of the individual electronic states within the band at different momenta *k* surrounding the K point can be plotted as a function of the magnetic field (Fig. 4e-g). We stress that the shift of all states shows a linear dependence following the Zeeman effect, but with *k*-dependent slopes. The states near the band bottom (*k*=0 referenced to the K point) exhibit the most pronounced evolution with energy (Fig. 4e). As the momentum moves away from the Dirac band bottom, the energy shift velocity gradually decreases, but saturates at a constant value of 0.48 meV/T (Fig. 4f-h). This momentum-dependent band shift is remarkable and has not been experimentally observed before. It directly reveals for the first time a *k*-dependent *g*-factor for Bloch electrons. Writing the energy shift as: $\Delta E(k) = -\frac{1}{2} g_k \mu_B B$, where *B* is the magnitude of the magnetic field, we extract the *k*-dependent *g*-factor $g_k$ and plot them as a function of *k* in Fig. 4j. Clearly, the behavior of the *g*-factor follows $g_k = g_0 + \delta g_k$, describing a constant background contribution $|g_0| \approx$ 15 and a *k*-dependent Lorentzian-like $|\delta g_k|$ ranging from 0 to 12, peaking around the K point at *k*=0.

Theoretically, there are several contributions that can in principle lead to the *k*-dependent shift of Dirac states most pronounced near the bottom of the observed electron-like part of the massive Dirac bands. First, note that the Dirac mass gap is present already at zero magnetic field. Similarly to the momentum-dependent *g*-factor in certain graphene-based structures [8–10], non-trivial Berry curvature associated with the massive Dirac fermions generates an orbital magnetic moment, and the orbital Zeeman coupling to the magnetic field along the c-axis is consistent with the observed energy shift linear in magnetic field (Zeeman shift). Our calculation in a simplified model of this system, based on a single-orbital, spin-polarized kagome lattice model confirms the large orbital magnetic moments, primarily localized near the Dirac points at K where the Berry curvature concentrates (Supplementary Information 5, Fig. S9c). The coupling to the applied magnetic field can account for the shift in the energy of the states linearly by orbital Zeeman effect, in a *k*-dependent fashion set by *k*-dependent orbital moments. Second, as magnetization measurements of $YMn_6Sn_6$ indicate canting of spins towards the field direction (Fig. S10), the Kane-Mele type SOC due to the in-plane orbital motion can naturally enhance the Dirac gap as the magnetization gains an out-of-plane component (Supplementary Section 6) [47]. It is thus expected that the observed rounding of the Dirac band bottom (Fig. 4c) contains contribution from the SOC as the out-of-plane magnetic field is applied. Lastly, we note that magnetic exchange coupling, as well as the intricate spin-textures [41,43] and its evolutions under a c-axis field currently under intensive investigation, can in principle also lead to additional field-induced band renormalization [48]. Future experiments probing the evolution of electronic properties under a vector magnetic field may provide further insights necessary to disentangle different underlying contributions to the observed momentum-dependent $g$-factor and fully understand the magnetic field tuning of the Dirac states revealed here.

Transport measurements have reported unconventional responses to magnetic field in a variety of bulk single crystals [49–51], but a direct insight into the field-induced electronic band structure change has been challenging. Our experiments reveal a momentum-dependent effective $g$-factor associated with a massive Dirac band in a kagome magnet. We note that although previous experiments reported a rigid field-induced shift of the spectral peaks associated with flat portions of the kagome bands [11,52], the bottom of the Dirac band [39], or localized states [53,54], our work for the first time reveals the continuous changes in associated electronic momenta. We note that although the origin of the Dirac mass at zero field is beyond the scope of this work, it is conceivable that the SOC in $YMn_6Sn_6$ stems beyond the Kane-Mele scenario and requires the full crystal symmetry and both in-plane and out-of-plane spin components. In addition, the spin Berry phase due to noncoplanar spin textures may also contribute to the zero-field Dirac mass, as in the case of $Fe_3Sn_2$ [39]. Due to the close proximity of the Dirac points to Fermi level in $YMn_6Sn_6$, the anomalous physics reported here could also be potentially probed by transport experiments. Given the complexity of the bulk electronic and magnetic structures of this system composed of at least eight kagome layers within a magnetic unit cell [43], a complete quantitative understanding of the momentum dependent $g$-factor will likely require a thorough many-body theoretical calculation, taking into account the helical magnetic structure and contributions from other Dirac and flat bands, including exchange effects [48]. As the $k$-dependent $g$-factor can produce a large band renormalization, it in turn enables fine tunability of the electronic structure by external magnetic field, including the possibility of driving the many-body ground states through different topological phase transitions. Since the magnetic order at the surface of $YMn_6Sn_6$ has broken the time-reversal symmetry, the gapped Dirac band should carry a nonzero Chern number, pointing to $YMn_6Sn_6$ being a topological kagome lattice Chern magnet.

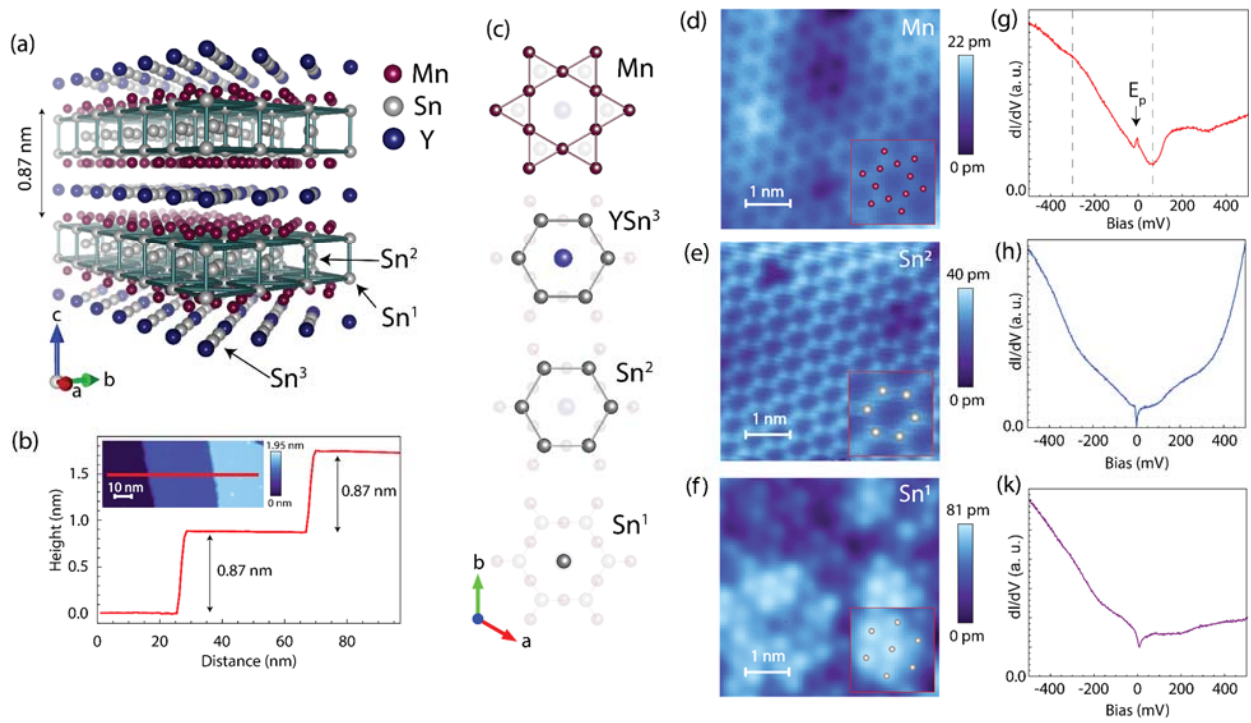

**Figure 1. Crystal structure and different surface terminations**. (a) The schematic of $YMn_6Sn_6$ crystal structure. (b) A topographic line profile across consecutive terraces, extracted along the red line in the inset. Each step shows a single unit cell step height of ~0.87 nm. (c) Different possible surface terminations associated with the *ab*-plane: Mn, $YSn^3$, $Sn^2$, and $Sn^1$ layers. (d-f) STM topographs and (g-k) average dI/dV spectra on: (d,g) the kagome Mn layer, (e,h) $Sn^2$ layer and (f,k) $Sn^1$ layer. A spectral peak labeled as $E_p$ is present near Fermi level on the Mn layer, but not on the $Sn^1$ or the $Sn^2$ layer, which show a small "dip" at the Fermi level often observed on non-kagome terminations in related systems [33,46]. Dashed lines in (g) denote the approximate energy range of linearly dispersing differential conductance. STM setup condition: (b) $I_{set}$ = 10 pA, $V_{sample}$ = 1 V, (d-f) $I_{set}$ = 70 pA, $V_{sample}$ = 30 mV. (g-k) $I_{set}$ = 300 pA, $V_{sample}$ = 500 mV, $V_{exc}$ = 2 mV.

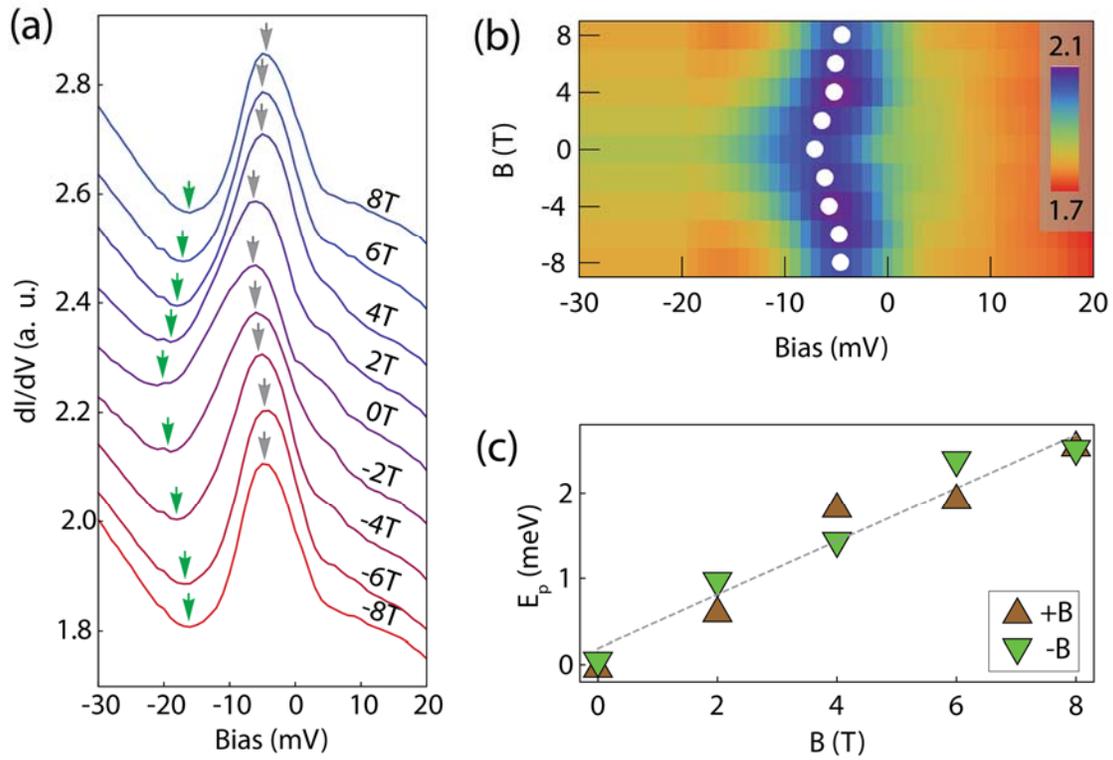

**Figure 2. Magnetic field dependence of the spectral peak in the density-of-states.** (a) Average d$I$/d$V$ spectra acquired over the identical region of the Mn layer from -8 T (antiparallel to *c*-axis) to 8 T (parallel to *c*-axis). The green and gray arrows indicate the approximate positions of the dip and the peak at $E_p$, respectively. (b) d$I$/d$V$ spectra from panel (a) in a different color scale for visualization purposes. White circles in (b) denote peak positions extracted by fitting (Fig. S5). (c) The scatterplot of $E_p$ as a function of applied magnetic field, showing linear dispersion with slope α = 0.31 ± 0.03 meV/T. STM setup condition: $I_{set}$ = 70 pA, $V_{sample}$ = 30 mV, $V_{exc}$ = 1 mV.

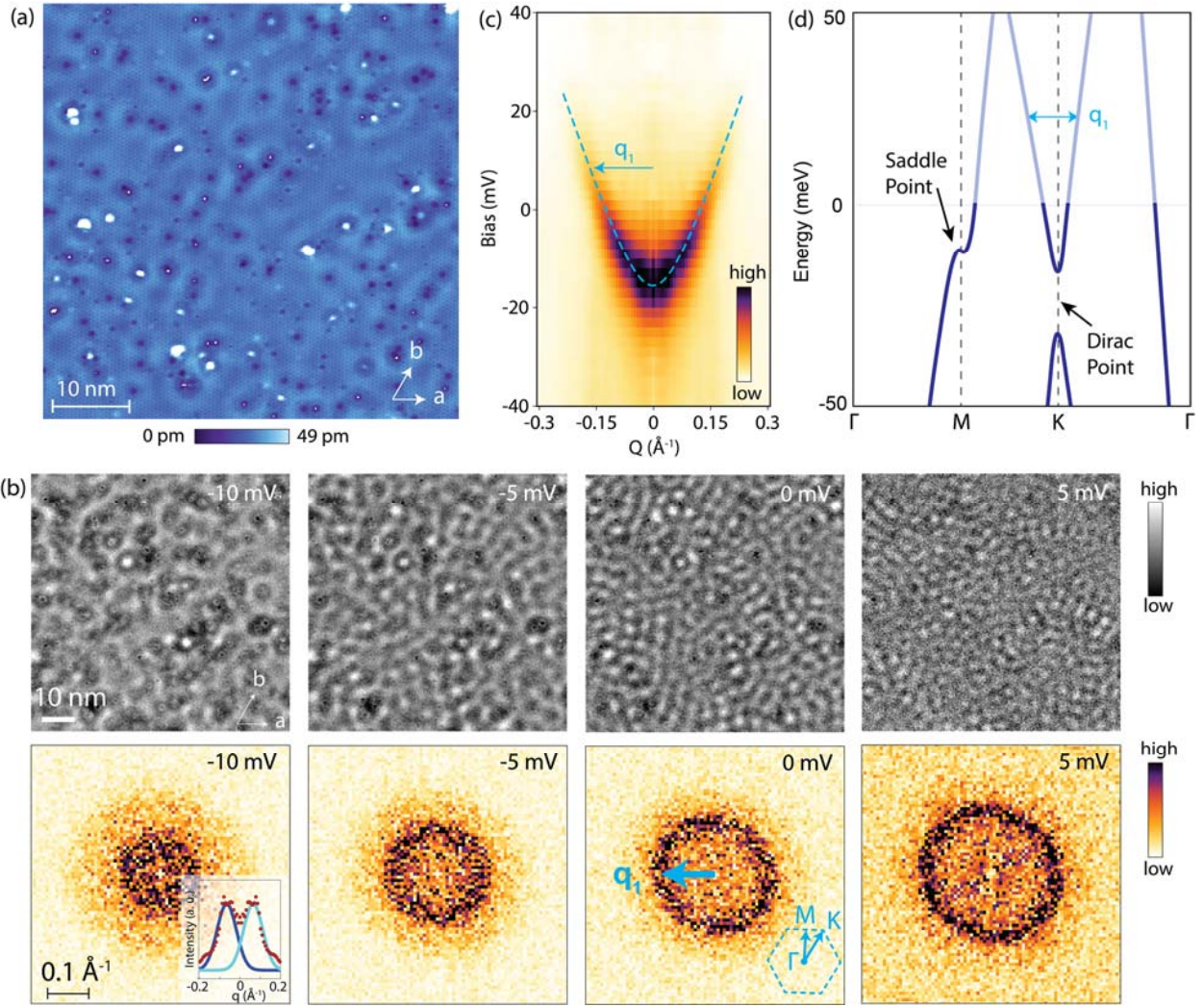

**Figure 3. Spectroscopic imaging of the Mn plane.** (a) Atomically-resolved STM topograph of a large, flat region partially encompassing the area where dI/dV maps in (b) are taken. (b) dI/dV maps over an identical region of the sample (top row) and their associated Fourier transforms (FTs) showing scattering of electrons on the surface (bottom row). Inset in (b) in the FT at -10 mV shows the radially averaged linecut (red points are the data) with clearly distinguishable scattering peaks (two blue curves are Gaussian fits). (c) Radially averaged linecut in FTs of dI/dV maps. Label $q_1$ in (b,c) denotes the prominent scattering vector in momentum-transfer space, which shows nearly isotropic spatial signature and a rapid evolution with energy. The dispersion velocity of the band associated with QPI signal in (c) is calculated to be 373.6 ± 9.4 meV·Å. The center pixel of the Fourier transform linecut in (c) has been artificially suppressed to emphasize other features. (d) Schematic of the band structure based on angle-resolved photoemission spectroscopy (ARPES) [45], with the Dirac point and the saddle point called out by arrows. Fermi level was shifted down by about 30-40 meV compared to ARPES measurement to match the energy of STM spectral features. STM setup condition: (b) $I_{set}$ = 70 pA, $V_{sample}$ = 30 mV, $V_{exc}$ = 2 mV, B = 0 T; (c) $I_{set}$ = 140 pA, $V_{sample}$ = 100 mV, $V_{exc}$ = 4 mV, B = 0 T.

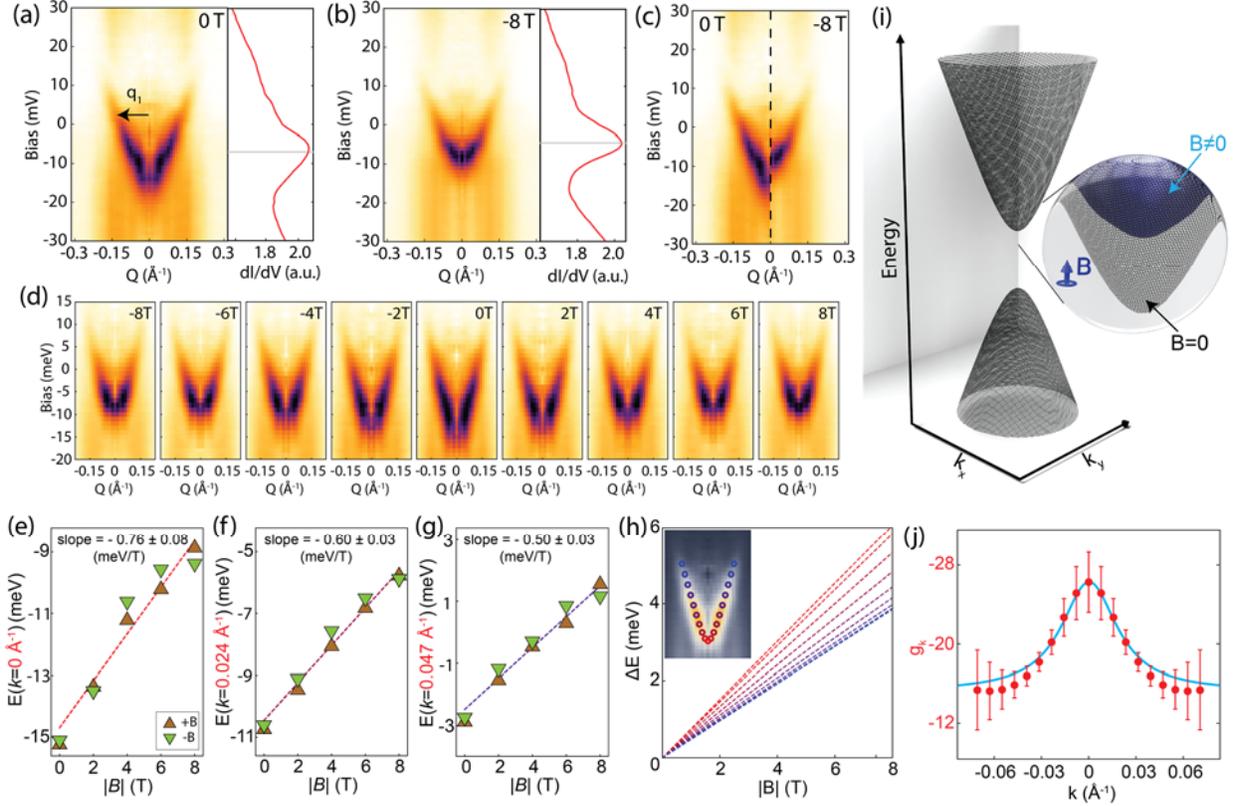

**Figure 4. Determination of the *k*-dependent *g*-factor.** (a,b) Radially averaged linecuts in Fourier transforms (FTs) of dI/dV maps at (a) 0 T and (b) -8 T. The right sides of panels in (a,b) show the corresponding average dI/dV spectra. (c) A direct comparison between radially-averaged Fourier space linecuts at 0 T (left) and -8 T (right). (d) A series of radially-averaged FT linecuts taken over the same region of the sample at different magnetic fields applied parallel and antiparallel to *c*-axis. (e-g) Examples of the dispersion of individual quantum states at (e) $k = 0$, (f) $k = 0.024$ Å$^{-1}$ and (g) $k = 0.047$ Å$^{-1}$ with magnetic field. (h) Linear fits to the data for a wide range of *k*'s showing a gradual change in slope, saturating at about - 0.48 meV/T slope (blue dashed line). Inset in (h) shows the color-coded momenta associated with different linear fits in the main panel. (i) Schematic of the Dirac band evolution in magnetic field. Blue (gray mesh) cones in the zoom-in depict the massive Dirac fermion at non-zero (zero) external magnetic field. (j) The measured *g*-factor $g_k$ for different electronic states in momentum space. Each point in (j) was obtained using the same extraction procedure shown in (e-h). Error bars in (j) represent two standard deviations obtained from the linear fits in ΔE *vs* B dispersions in (e-h). We also apply 1:2 conversion between momentum-transfer space (*q*-space) and momentum space (*k*-space) in (e-h,j). STM setup condition: $I_{set}$ = 70 pA, $V_{sample}$ = 30 mV, $V_{exc}$ = 1 mV.

(2020).

## Methods

**Sample growth and characterizations.** Single crystals of YMn$_6$Sn$_6$ were grown by using Sn flux. Y lumps (purity 99.99 %), Mn granules (purity 99.9 %) and Sn grains (purity 99.99 %) with a molar ratio of Y : Mn : Sn = 1 : 6 : 30 were put into an alumina crucible and sealed in a quartz ampoule under partial argon atmosphere. The sealed quartz ampoule was heated up to 1273 K and held for 24 hours. Then it was cooled down slowly to 873 K at a rate of 5 K/hours. Finally, the ampoule was taken out from the furnace and decanted with a centrifuge to separate YMn$_6$Sn$_6$ crystals from excess Sn flux.

**STM experiments.** Single crystals of YMn$_6$Sn$_6$ were cleaved in ultra-high vacuum, and immediately inserted into STM head where they are held at about 4.5 K during the measurements. Out of more than 40 crystals for which cleaving was attempted, we successfully cleaved and approached on 7 different samples. Two of them showed large enough Mn terraces for high-resolution spectroscopic mapping (Fig. 4 and Fig. S11). STM data was acquired using a customized Unisoku USM1300 STM. Spectroscopic measurements were made using a standard lock-in technique with 915 Hz frequency and bias excitation as detailed in figure captions. Zero energy of our measurement was checked, and no noticeable difference was observed as a function of magnetic field (Fig. S8). STM tips used were home-made chemically-etched tungsten tips, annealed in UHV to bright orange color prior to STM experiments.

## Competing Interests

The Authors declare no Competing Financial or Non-Financial Interests.

## Code availability

The computer code used for data analysis is available upon request from the corresponding author.

## Data Availability

The data supporting the findings of this study are available upon request from the corresponding author.

## Acknowledgements

I.Z. gratefully acknowledges the support from Army Research Office Grant No. W911NF-17-1-0399. H.C.L. acknowledges support by the National Key R&D Program of China (Grant No. 2018YFE0202600 and 2016YFA0300504), the National Natural Science Foundation of China (Grant No. 11774423 and 11822412), the Beijing Natural Science Foundation (Grant No. Z200005), and the Fundamental Research Funds for the Central Universities, and the Research Funds of Renmin University of China (18XNLG14, and 19XNLG17). Theoretical work at Renmin University was supported by the National Key R&D Program of China (Grant No. 2017YFA0302903), the National Natural Science Foundation of China (Grant No. 11774424), the Beijing Natural Science Foundation (Grant No. Z200005), the CAS Interdisciplinary Innovation Team, the Fundamental Research Funds for the Central Universities, and the Research Funds of Renmin University of China (Grant No. 19XNLG03). K.L. acknowledges the use of computational resources provided by the Physical Laboratory of High Performance Computing at Renmin University of China. Z.W. acknowledges the support from U.S. Department of Energy, Basic Energy Sciences Grant No. DE-FG02-99ER45747.